\documentclass[a4paper,twoside]{article}
\usepackage{epsfig}
\usepackage{subfigure}
\usepackage{calc}
\usepackage{amssymb}
\usepackage{amstext}
\usepackage{amsmath}
\usepackage{amsthm}
\usepackage{multicol}
\usepackage{pslatex}
\usepackage{apalike}
\usepackage{SCITEPRESS}
\usepackage[small]{caption}

\begin{document}

\title{Precise Phase Measurements using an Entangled Coherent State}

\author{\authorname{P.A. Knott\sup{1} and J.A. Dunningham\sup{1,2}} 
\affiliation{\sup{1}School of Physics and Astronomy, University of Leeds, Leeds LS2 9JT, United Kingdom}
\affiliation{\sup{2}Department of Physics and Astronomy, University of Sussex, Falmer, Brighton BN1 9QH, United Kingdom}
\email{phy5pak@leeds.ac.uk, j.dunningham@sussex.ac.uk}
}


\abstract{Quantum entanglement offers the possibility of making measurements  beyond the classical limit, however some issues still need to be overcome before it can be applied in realistic lossy systems. Recent work has used quantum Fisher information (QFI) to show that entangled coherent states (ECSs) may be useful for this purpose as they combine sub-classical phase precision capabilities  with robustness \cite{joo2011quantum}. However, to date no effective scheme for measuring a phase in lossy systems using an ECS has been devised. Here we present a scheme that does just this. We show how one could measure a phase to a precision significantly better than that attainable by both unentangled `classical' states and highly-entangled NOON states over a wide range of different losses. This brings quantum metrology closer to being a realistic and practical technology.}

\onecolumn \maketitle \normalsize \vfill

\section{\uppercase{Introduction}}
\label{sec:introduction}

Quantum metrology is the art of making precision measurements by taking advantage of the properties of quantum mechanics. The main advantage of quantum metrology over classical metrology is that it allows us to achieve the same precision with fewer resources. Making more precise measurements with limited numbers of particles or photons has many important applications, including microscopy, gravitational wave detection, measurements of material properties, and medical and biological sensing \cite{nagata2007beating}. Many of these examples could benefit  from a device which operates with a lower photon flux, for example in biological sensing, where disturbing the system too much can damage the sample. Another important reason for developing quantum metrology is that it provides a stepping stone to more complicated quantum technologies such as quantum computers: if we can build metrological devices that can beat the classical limit by exploiting entanglement, then this puts us in a good position to begin to tackle more advanced manipulations of quantum properties. Furthermore, measurements are crucial to the development of science, and any way to improve them is a welcome development.

One of the main stumbling points in quantum metrology is creating a state that is robust to particle losses, which will always be a concern in any realistic device. In this paper we present an optical state that shows remarkable robustness to photon losses: entangled coherent states ECSs. We discuss how these states can be created, and present a scheme which allows us to measure a phase using an ECS to sub-classical precision, even when a wide range of different loss rates are accounted for.

\section{\uppercase{Precise Phase Measurements}}
\label{sec:prec_phase_measurement}

\subsection{Enhancement Using Entanglement}
\label{sec:metrology}

\begin{figure}[!h]
  \vspace{-0.2cm}
  \centering
   {\epsfig{file = 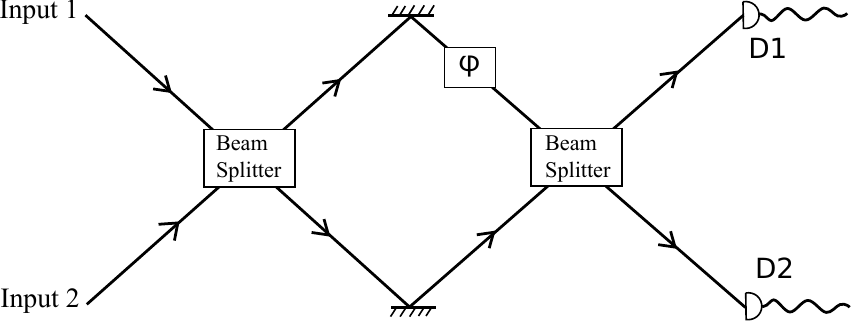, width = 7.5cm}}
  \caption{A Mach-Zehnder interferometer which can be used to measure a phase $\phi$. Photons are send through the beam splitters and phase shift, then measured at the detectors $D1$ and $D2$.}
  \label{fig:MZ_intf}
  \vspace{-0.1cm}
\end{figure}

Throughout this paper we will generally be concerned with measuring phases in a device using the same principles of a Mach-Zehnder interferometer, as shown in Fig.~\ref{fig:MZ_intf}. The first step in this device is to combine the two input states at a beam splitter. One of the paths then picks up a phase, $ \phi $, and  the two paths are recombined at a second beam splitter. The resulting output from the second beam splitter can be measured by detectors $D1$ and $D2$ to extract the phase information. A `classical-like' state of light, which is equivalent to sending a sequence of independent single particles (SP) through the interferometer allows one to measure the phase with a precision that scales with the total number of particles $n$ as the ``shot noise limit" $ 1/{\sqrt{n}}$ \cite{gkortsilas2012measuring}. However by making use of an entangled state, the precision can be improved to $ 1/{n}$: the ``Heisenberg limit" \cite{dunningham2006using}.
\\

In order to create an entangled number state we replace the first beam splitter in the interferometer in Fig.~\ref{fig:MZ_intf} by a ``quantum beam splitter" (QBS). A quantum beam splitter is like an ordinary interferometer, but with a nonlinearity in one arm \cite{dunningham2006using} instead of the phase, and has the following effect on state $|n,0 \rangle$:

\begin{align}
\label{eq:qbs2}
|n,0 \rangle \xrightarrow{\rm QBS} \frac{1}{\sqrt 2} \left[ |n,0\rangle + i |0,n\rangle \right]
\end{align}

The state on the right hand side is known as a NOON state \cite{dowling2008quantum}. After creating the NOON state, we then apply the phase shift to one of the paths in the interferometer, as we did with the SP state. We can then send this state through a second QBS, and measure the number of particles at the detectors. If we send a NOON state of $n$ particles through through this scheme, then the precision of phase measurement can be seen to vary as $\delta\phi = 1/n$, the Heisenberg limit \cite{dunningham2006using}.

However, there is a problem with such an approach because NOON states are highly fragile to particle loss. Losing just one particle from a NOON state (Eq.~ \ref{eq:qbs2}) will project the state onto either $e^{in\phi} |n,0\rangle$ or $i |0,n\rangle$. The global phase in each of these cases is not physical and cannot be measured, so all the phase information is lost when we lose just one particle. Despite this, a number of clever schemes have been devised with robustness to loss which still capture sub shot noise limit precision, albeit not quite at the Heisenberg limit. An example of one of these schemes is a NOON ``chopping" strategy \cite{dorner2009optimal}, in which multiple smaller NOONs are sent through an interferometer instead of one big one. Other examples include unbalanced NOON states \cite{demkowicz2009quantum}, BAT states \cite{gerrits2010generation} and mixtures of SP and NOON states \cite{gkortsilas2012measuring}. While these states can beat the shot noise limit when loss is included, they are still fragile, and with large amounts of loss they are outperformed by classical strategies. We will now turn to a state which shows huge potential, as it is intrinsically robust to the effects of loss: coherent states.

\subsection{Entangled Coherent States}

A coherent state is defined as:

\begin{eqnarray}
|\alpha \rangle = e^{-\frac{|\alpha|^2}{2}}\sum_{n=0}^{\infty}\frac{\alpha^n}{\sqrt{n!}}|n\rangle,
\end{eqnarray}
where $\alpha$ is a complex amplitude and $n$ is the particle number. In order to achieve quantum enhanced measurement, we still need to create an \textit{entangled} coherent state (ECS) \cite{sanders2012review,hirota2011effectiveness}:

\begin{eqnarray}
\label{eqn:ECS_general}
\mathcal{N} \left( |\alpha,0 \rangle + e^{i\theta} |0,\alpha \rangle \right)
\end{eqnarray}
where $\mathcal{N}=1/ \sqrt{2+2e^{-|\alpha|^2}\cos{\theta}}$. We could create the ECS with $\theta=\pi /2$ by sending input state $|\alpha,0 \rangle$ through the QBS. Alternatively, the ECS with $\theta=0$ can be created by interacting a ``cat state'' $N_{\alpha} (|+\alpha \rangle + |-\alpha \rangle)$ with a coherent state $|\alpha \rangle$ at a beam splitter \cite{joo2011quantum}. This latter scheme is likely to be more experimentally feasible. However, the issue of experimental implementations will be left to later work.

In order to investigate the potential of the ECS in quantum metrology, Joo \textit{et. al.} \cite{joo2011quantum} calculated its quantum Fisher information (QFI). The QFI for a general state $\rho$ \cite{braunstein1994statistical,boixo2009quantum} is given by:

\begin{eqnarray}
F_Q = \text{Tr} \left( \rho A^2 \right)
\end{eqnarray}
where $A$ is found from solving the symmetric logarithmic derivative $\partial\rho/\partial\phi = 1/2 \left[ A \rho + \rho A \right]$. The precision in the phase measurement (more specifically the lower bound on the standard deviation) is given by the quantum Cram\'er-Rao bound \cite{braunstein1994statistical}:
\begin{eqnarray}
\delta \phi \ge {1 \over \sqrt{\mu F_Q}},
\end{eqnarray}
where $\mu$ is the number of copies of the state (i.e. times that the measurement is independently repeated).
This gives the best possible precision with which a state can measure a phase. For NOON states and SP states the quantum Cram\'er-Rao bound gives us the Heisenberg and shot noise limits respectively.

Joo \textit{et. al.} used the QFI to show that with and without loss the ECS can achieve better precisions than SP, NOON, and some other candidate states. Zhang \textit{et. al.} \cite{zhang2013quantum} derived an expression for the QFI with loss for arbitrary $\alpha$, and confirmed the potential of ECSs for robust quantum metrology. What is still missing, however, is a concrete way for converting this promise into a real scheme for making the phase measurement.

\section{\uppercase{Measuring a phase using an entangled coherent state}}
\subsection{A Simple Scheme with No Loss}
\label{sec:no_loss_simple}

We will now look at the effect of sending input state $|\psi_1 \rangle = |\alpha,0 \rangle$ through the interferometer in Fig.~\ref{fig:MZ_intf}, but with the beam splitters replaced by QBSs. The effect of the first QBS is to produce the ECS $|\psi_2 \rangle$:

\begin{align}
|\psi_1 \rangle = |\alpha,0 \rangle &\xrightarrow{\rm QBS} {1 \over \sqrt{2}} \left( |\alpha,0 \rangle + i |0,\alpha \rangle \right) = |\psi_2 \rangle.
\end{align}
We then perform a phase shift which gives us the state:

\begin{align}
\label{eqn:ECS_phased}
|\psi_3 \rangle \notag&= \frac{e^{-\frac{|\alpha|^2}{2}}}{\sqrt{2}} \sum_{n=0}^{\infty}\frac{\alpha^n}{\sqrt{n!}} \left[ e^{in\phi}|n,0\rangle + i |0,n\rangle \right] \\ &= \frac{1}{\sqrt{2}} \left( |\alpha e^{i\phi},0 \rangle + i|0, \alpha \rangle \right),
\end{align}
followed by the second QBS:

\begin{eqnarray*}
\label{eq:phase_measure}
|\psi_4 \rangle = e^{-\frac{|\alpha|^2}{2}} \sum_{n=0}^{\infty}\frac{\alpha^n}{\sqrt{n!}} ie^{in\phi \over 2} \left(|n,0\rangle \sin{n\phi \over 2} + |0,n\rangle \cos{n\phi \over 2} \right).
\end{eqnarray*}

From this we can calculate the probability amplitude of detecting different numbers of photons at the outputs. To do this we first take the inner product of $|\psi_4 \rangle$ with $ | n_1 \rangle_{_{D1}} |n_2 \rangle_{_{D2}} = | n_1,n_2 \rangle$, i.e. the state with $n_1$ photons at detector $D1$ and $n_2$ photons at detector $D2$. This gives us:

\begin{align*}
\label{}
&\langle n_1,n_2 |\psi_4 \rangle = \\ &ie^{-\frac{|\alpha|^2}{2}} \left[ \frac{\alpha^{n_1}}{\sqrt{n_1!}} e^{in_1\phi \over 2} \sin{n_1\phi \over 2} \delta_{n_2,0} + \frac{\alpha^{n_2}}{\sqrt{n_2!}} e^{in_2\phi \over 2} \cos{n_2\phi \over 2} \delta_{n_1,0} \right].
\end{align*}

The delta functions here tell us that it is impossible to detect photons at both outputs. This is clearly true as any photon detection collapses the state into either $|n,0 \rangle$ or $ |0,n \rangle$. We can now calculate the probabilities of different numbers of photons being detected, given that the phase in the interferometer is $\phi_1$:

\begin{eqnarray}   
\label{eq:Prob_phi_given}
P(n_1,n_2|\phi=\phi_1) = \left| \langle n_1,n_2 |\psi_4 \rangle \right|^2
\end{eqnarray}

\begin{figure}[!h]
  \vspace{-0.2cm}
  \centering
   {\epsfig{file = 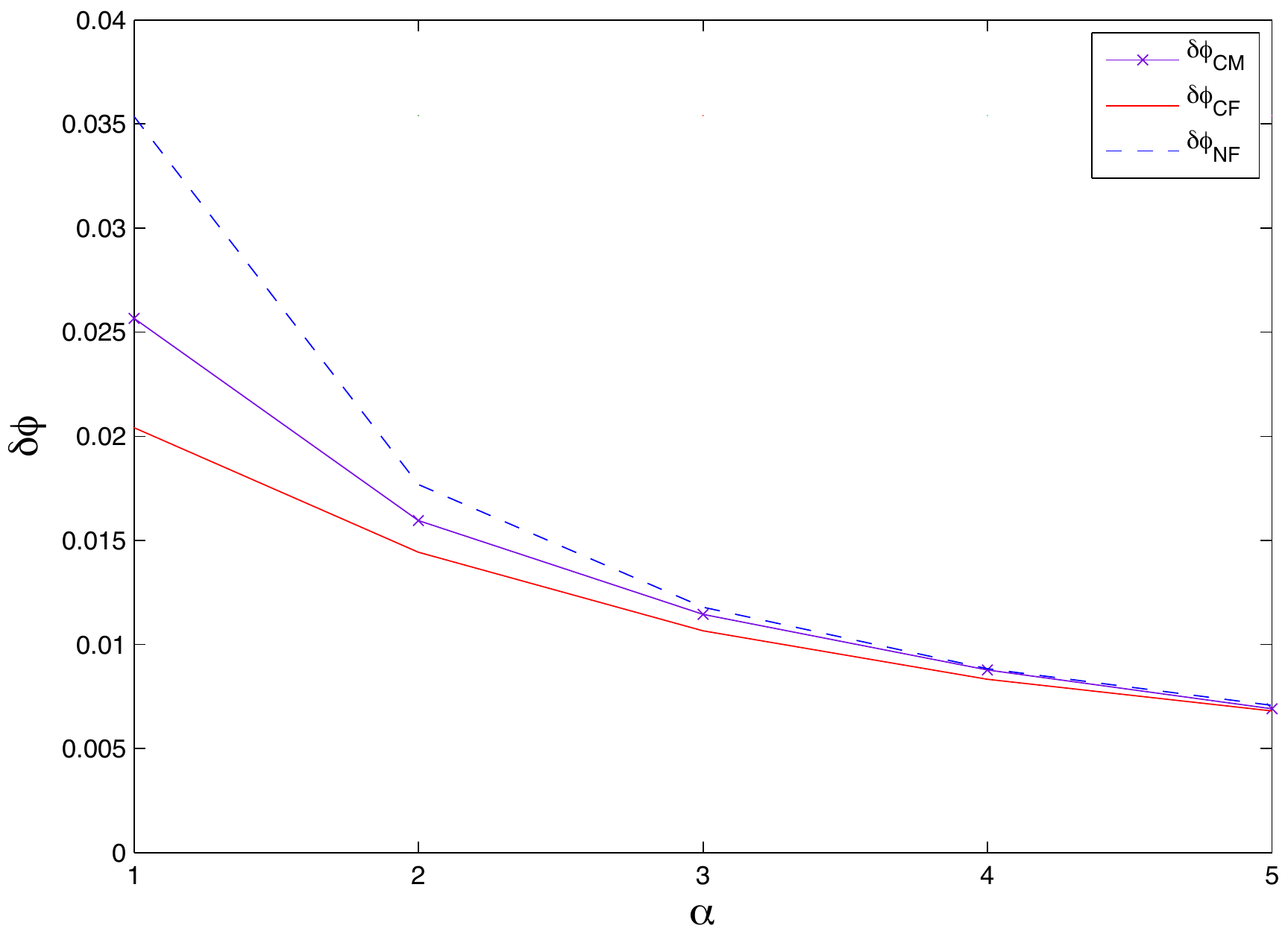, width = 7.5cm}}
  \caption{The phase precision of the scheme described in section \ref{sec:no_loss_simple} for no loss is shown here. We plot different sized ECSs against their precision. Here (and in later figures) $\delta\phi_{CM}$ is the ECS using our measurement scheme, $\delta\phi_{CF}$ is the QFI for the ECS and $\delta\phi_{NF}$ is the QFI for the NOON state (of equivalent size as each ECS). SP states measure at precision 0.0354.}
  \label{fig:alpha_vs_prec}
  \vspace{-0.1cm}
\end{figure}

Using this conditional probability distribution we can apply Bayesian statistics to build up our knowledge of the phase $\phi$ as we repeat the process with a stream of ECSs \cite{gkortsilas2012measuring}. Fig.~\ref{fig:alpha_vs_prec} shows that this scheme, with no loss, allows us to beat the best possible precision obtainable using NOON states of comparable sizes. For small $\alpha$ we do not saturate the QFI, but we significantly improve upon the best possible measurement using a NOON state. For large $\alpha$ this scheme comes very close to saturating the QFI, but it can be seen that in this region ECSs and NOON states operate at a very similar precision.

We will now briefly discuss the details of our scheme that allow us to compare our results to NOON and SP states. In many applications of quantum metrology such as biological sensing, probing materials’ properties and gravitational wave detection, we are concerned with the number of photons (or particles) passing through the phase shift itself, and would like to minimise this number when possible. We would therefore like to compare the phase measurement precision of different states when a set number of photons $R$ enter the phase. For the unentangled state we simply send the state through the interferometer $2R$ times, as in each run an average of $1/2$ a photon enters the phase shift. For the NOON state in equation \ref{eq:qbs2}, each run sends $n/2$ photons through the phase shift, and so we simply send the NOON state through the interferometer $2R/n$ times. For ECSs the situation is slightly different, as each run contains a different number of photons. We can calculate the average number of photons passing through the phase for the general ECS in equation \ref{eqn:ECS_general} to be $\bar{n}=\mathcal{N}^2 |\alpha|^2$, and we therefore send the ECS through the interferometer $R/\bar{n}$ times. For most of our results we have used $R=400$ as this allows us to consistently calculate the precision of the phase measurement with different states.

\subsection{Introducing Loss}
\label{sec:loss_more_realistic}

\begin{figure}[!h]
  \vspace{-0.2cm}
  \centering
   {\epsfig{file = 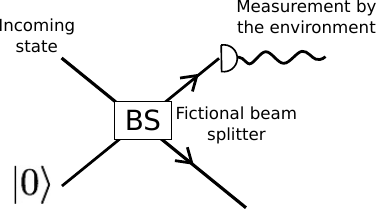, width = 5.5cm}}
   \caption{This shows a ``fictional'' beam splitter and measurement by the environment as a model for loss.}
  \label{fig:ECS_loss_FBS}
  \vspace{-0.1cm}
\end{figure}

To simulate the effects of loss we introduce ``fictional'' beam splitters after the phase shift \cite{gkortsilas2012measuring,joo2011quantum,demkowicz2009quantum} as shown in Fig.~\ref{fig:ECS_loss_FBS}, which have probability of transmission $\eta\equiv \cos^{2}\theta$. After the phase shift and including the vacuum states used to simulate loss, the ECS we are concerned with is given by:

\begin{eqnarray}
\label{}
|\psi_0 \rangle = \frac{1}{\sqrt{2}} \left[ |\alpha e^{i\phi},0,0,0\rangle + i |0,0,\alpha ,0 \rangle \right],
\end{eqnarray}
where the second and fourth terms in the kets represent the modes into which particles are lost from the first and third terms respectively.

The effect of the ``fictional'' beam splitters then leaves us in the state $|\Psi_1 \rangle$ given by:

\begin{align*}
\label{}
\frac{1}{\sqrt{2}} \left[ |\alpha e^{i\phi} \sin{\theta},\alpha e^{i\phi} \cos{\theta},0,0\rangle + i |0,0,\alpha \sin{\theta},\alpha \cos{\theta} \rangle \right].
\end{align*}

We then take the density matrix $\rho_1 = | \psi_1 \rangle \langle \psi_1 |$ and to represent measurement by the environment we trace over the environmental modes as follows:

\begin{eqnarray}
\label{}
\rho_2 = \sum_{e_1} \sum_{e_2} \langle e_1 | \langle e_2 | \rho_1 | e_2 \rangle | e_1 \rangle.
\end{eqnarray}

Using $ \sum_{e} \langle e | X \rangle \langle Y | e \rangle = \langle Y | X \rangle $ and the nonorthogonality of coherent states $ \langle \alpha | \beta \rangle = \exp{(-{1 \over 2} |\alpha|^2 + \alpha^* \beta - {1 \over 2} |\beta|^2)} $ it can be shown that $\rho_2 $ is reduced to:

\begin{align*}
\label{}
&\rho_2 = c_1 \left( | \psi_2 \rangle \langle \psi_2 | \right) + \\ &{1 \over 2} c_2 \left( |\alpha e^{i\phi} \sqrt{\eta},0 \rangle \langle \alpha e^{i\phi} \sqrt{\eta},0 | + |0,\alpha \sqrt{\eta} \rangle \langle 0,\alpha \sqrt{\eta} | \right)
\end{align*}

where $c_1=e^{|\alpha|^2 (\eta-1)}$, $c_2=1-c_1$ and:

\begin{eqnarray}
|\psi_2 \rangle = \frac{1}{\sqrt{2}} \left[ |\alpha e^{i\phi} \sqrt{\eta} ,0\rangle + i |0,\alpha \sqrt{\eta} \rangle \right].
\end{eqnarray}

\begin{figure}[!h]
  \vspace{-0.2cm}
  \centering
   {\epsfig{file = 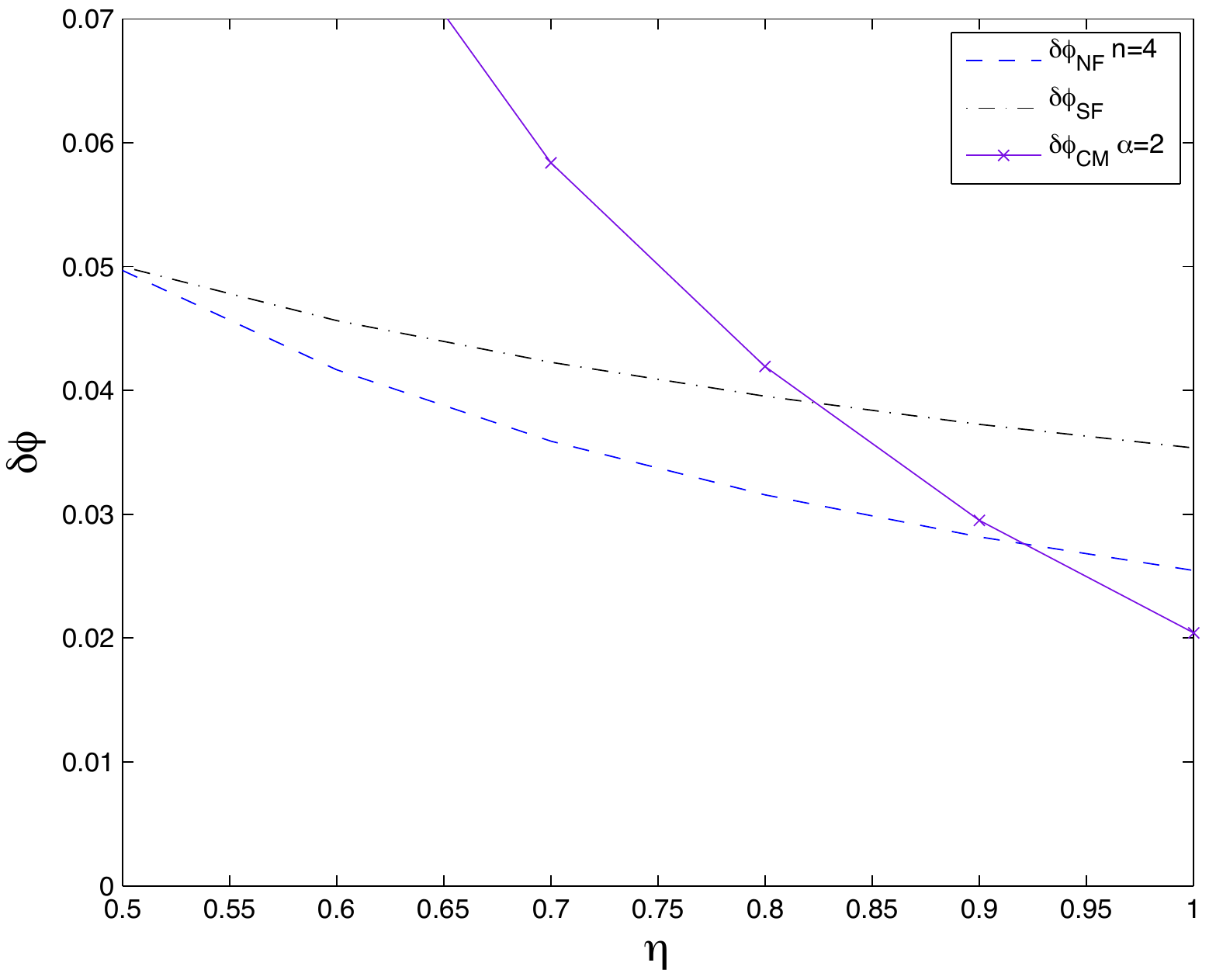, width = 7.5cm}}
  \caption{Here the legend refers to states as in Fig.~\ref{fig:alpha_vs_prec}, with $\delta\phi_{SF}$ the QFI for SP states. We can see that ECSs degrade quickly with loss ($\alpha=\sqrt{2}$ here). For larger $\alpha$ the ECS loses precision with loss even quicker.}
  \label{fig:simple_loss}
  \vspace{-0.1cm}
\end{figure}

The resulting state is a mixture of loss and no loss components. For NOON states we know during each run if there has been loss simply by counting the numbers of particles at the outputs. However, for ECSs we can no longer do this, as we don't know the number of particles in an ECS to begin with. This, combined with the fact that the size of the coherent state has decreased after the loss, means that the ECS in this simple interferometer loses its phase precision more quickly than NOON states, as shown in Fig.~\ref{fig:simple_loss}. This agrees with the work by \cite{zhang2013quantum} who calculated the Fisher information of ECSs with loss for any value of $\alpha$. They showed that the quantum enhancement of the ECS decreases more quickly than NOON states with loss. Despite this they also showed that the ECSs still contain some phase information after loss (unlike NOON states) and this should allow us to recover the phase information and therefore end up beating NOON states in the long run. Our simple scheme clearly does not do this, but in the next section we will present a scheme that does.

\section{\uppercase{Improved Scheme with Loss}}
\label{sec:long_arm}

\begin{figure}[!h]
  \centering
   {\epsfig{file = 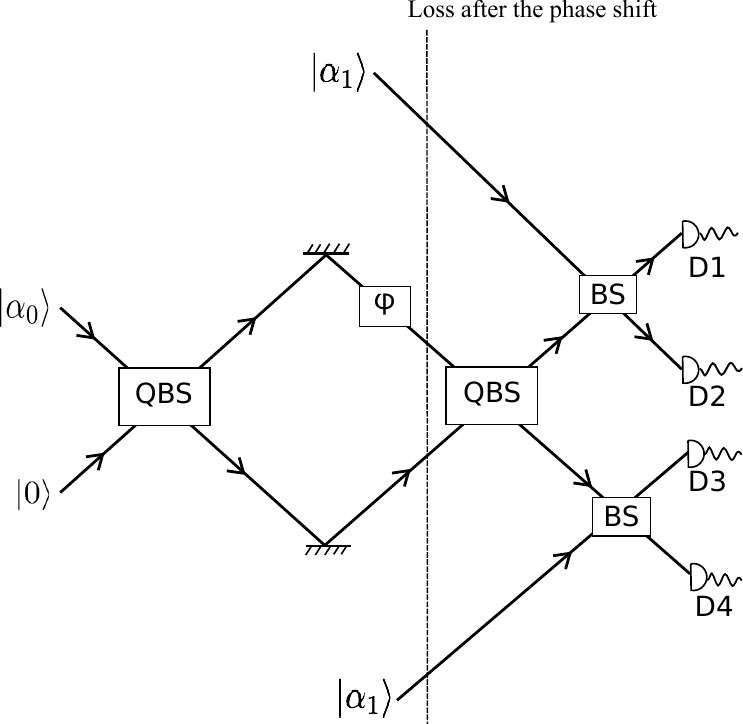, width = 8cm}}
  \caption{Quantum interferometer with extra arms to recover phase information with loss.}
  \label{fig:ECS_longarm2}
 \end{figure}

Despite the fact that an entangled coherent state can still retain some phase information after loss, we have seen that with a simple measurement scheme the phase information cannot be recovered, and we end up doing even worse than NOON states. We have devised a scheme, shown in Fig.~\ref{fig:ECS_longarm2}, which can be used to recover this desired phase information. The key is to use extra ``reference'' coherent states above and below the main interferometer which can be used to perform a homodyne measurement and recover the phase information. When a photon is lost from the ECS in equation \ref{eqn:ECS_phased} the state collapses into $|\alpha e^{i\phi},0 \rangle$ or $|0,\alpha \rangle$. If we are left with the second state, then the phase information is irretrievable, but if we are left with the state $|\alpha e^{i\phi},0 \rangle$ then the phase information is still there. However, in order to extract it we need a reference state $|\alpha \rangle$ to ``compare'' it to, hence including the upper and lower arms in our interferometer.

The state in this ``long arm'' interferometer after the phase shift is:

\begin{align}
\label{eqn:initial_VBS}
|\Psi_1 \rangle &= {1 \over \sqrt{2}} \left( |\alpha_1,\alpha_0 e^{i\phi},0,\alpha_1 \rangle + i|\alpha_1,0,\alpha_0,\alpha_1 \rangle \right) \\ &= |\Phi_1 \rangle + |\Phi_2 \rangle.
\end{align}

\begin{figure}[!h]
  \vspace{-0.2cm}
   {\epsfig{file = 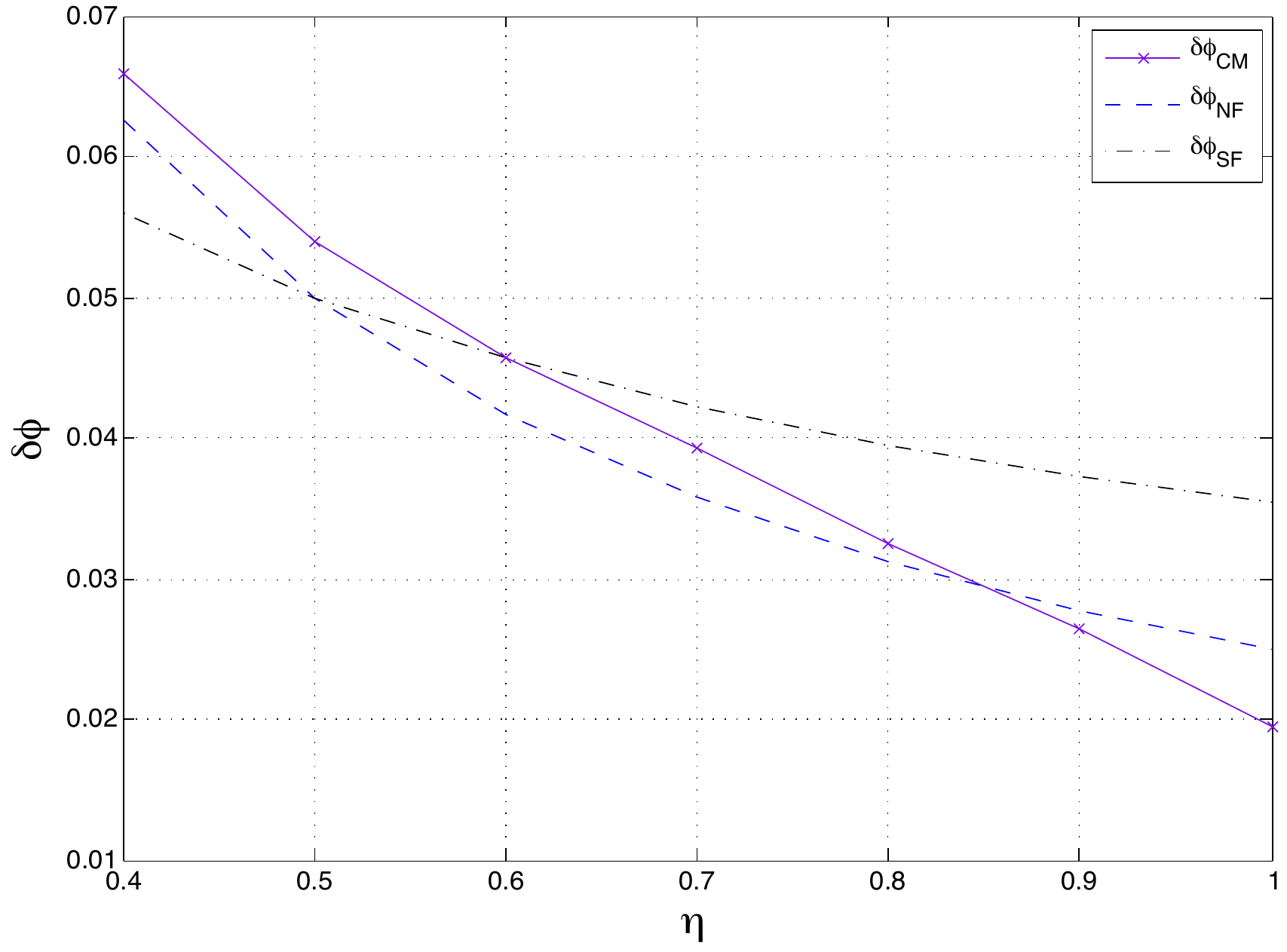, width = 7.5cm}}
  \caption{This scheme clearly doesn't perform well compared to NOON and SP when there is more than 15\% loss.}
  \label{fig:alrt2_optphi_cheta}
  \vspace{-0.1cm}
\end{figure}

After being acted on by the fictional beam splitters that simulate loss, this state is transformed from $|\Psi_1 \rangle$ to $|\Psi_2 \rangle$. We then trace over the environmental degrees of freedom to give $\rho_1 = \sum_{e}  \langle e_2 | \Psi_2 \rangle \langle \Psi_2 | e \rangle$ 	where $|e \rangle$ represents all four environmental modes. This gives us:

\begin{align}
\label{}
\rho_2 &= |\Phi_{1\eta} \rangle \langle \Phi_{1\eta} | + |\Phi_{2\eta} \rangle \langle \Phi_{2\eta} | \\ &+ e^{-|\alpha_{0\mu}|^{2}} \left( |\Phi_{1\eta} \rangle \langle \Phi_{2\eta} | + |\Phi_{2\eta} \rangle \langle \Phi_{1\eta} | \right),
\end{align}
where $\eta$ is the transmission rate through the interferometer, $|\Phi_{1\eta}\rangle={1 \over \sqrt{2}} |\alpha_{1\eta} , \alpha_{0\eta}e^{i\phi}, 0,  \alpha_{1\eta} \rangle$, $|\Phi_{2\eta}\rangle={i \over \sqrt{2}} |\alpha_{1\eta} , 0,  \alpha_{0\eta},  \alpha_{1\eta} \rangle$, $\alpha_{0\eta}=\alpha_0 \sqrt{\eta}$, $\alpha_{1\eta}=\alpha_1 \sqrt{\eta}$ and $\alpha_{0\mu}=\alpha_0 \sqrt{1-\eta}$. This state can also be written as:

\begin{eqnarray*}
\label{}
\rho_2 = c_1 |\Psi_{1\eta} \rangle \langle \Psi_{1\eta} | + c_2 \left( |\Phi_{1\eta} \rangle \langle \Phi_{1\eta} | + |\Phi_{2\eta} \rangle \langle \Phi_{2\eta} | \right)
\end{eqnarray*}
where $|\Psi_{1\eta} \rangle$ is the state in equation \ref{eqn:initial_VBS} with all the $\alpha$ reduced to $\sqrt{\eta}\alpha$. In this form it is easy to see that the mixed state after loss is comprised of a pure state $|\Psi_{1\eta} \rangle$ with coefficient $c_1=e^{-|\alpha_{0\mu}|^{2}}$, a collapsed state $|\Phi_{1\eta}\rangle$ which contains the phase, and a collapsed state $|\Phi_{2\eta}\rangle$ that does not contain the phase, both with coefficient $c_2=1-c_1$. We can then send $\rho_2$ through the remainder of the interferometer, giving the probabilities at the outputs as:

\begin{align*}
\label{}
P(\#) &= \langle \# |\overline{\Phi_{1\eta}} \rangle \langle \overline{\Phi_{1\eta}} | \# \rangle + \langle \# |\overline{\Phi_{2\eta}} \rangle \langle \overline{\Phi_{2\eta}} | \# \rangle \\ &+ e^{-|\alpha_{0\mu}|^{2}} \left[ \langle \# |\overline{\Phi_{1\eta}} \rangle \langle \overline{\Phi_{2\eta}} | \# \rangle + \langle \# |\overline{\Phi_{2\eta}} \rangle \langle \overline{\Phi_{1\eta}} | \# \rangle \right],
\end{align*}
where $|\#\rangle = |k,l,m,n\rangle$, the state with $k$ particles in the first output, $l$ in the second and so on. The barred states $|\overline{\Phi_{1\eta}}\rangle$ and $|\overline{\Phi_{2\eta}}\rangle$ can be found by sending $|{\Phi_{1\eta}}\rangle$ and $|{\Phi_{2\eta}}\rangle$ through the remainder of the interferometer. We initially took the obvious choice for the ``reference'' states as $\alpha_1=\alpha_0$. However, we found that this scheme gave us poor results, as shown in Fig.~\ref{fig:alrt2_optphi_cheta} where $\alpha=\sqrt{2}$. For this very small choice of $\alpha_0=\sqrt{2}$ we can beat the NOON and SP states up to around $15\%$ loss, which is indeed a very positive find. But after this point it is more beneficial to use either NOON or SP states. If we increase $\alpha$ then the results soon get much worse and before long we cannot beat either NOON or SP states if there is any loss at all.

Despite these shortcomings, the clear potential of ECSs warranted a more rigorous search for changes that can optimise our scheme. Indeed, if we instead use the initial ECS $\mathcal{N} ( |\alpha,0 \rangle + |0,\alpha \rangle ) $ where $\mathcal{N}=1/\sqrt{2(1+e^{-|\alpha|^2})}$, as used by Joo \textit{et. al.} \cite{joo2011quantum}, we begin to get more positive results. A much more significant change we can make is to vary the size of $\alpha_1$ for different loss values, in some cases up to around $\alpha_1=2.4\alpha_0$ -- a detailed study of why this is the case is the subject of ongoing work. The precision with which this scheme can measure the phase also depends on the (approximate) phase being measured (this is true for most schemes). Nonetheless this should not pose much of a problem as we can just put a variable phase shift in the lower-middle path, which allows us to vary the phase difference so that effectively $\phi$ can be whatever we choose.

\begin{figure}[!h]
  \vspace{-0.2cm}
  \centering
   {\epsfig{file = 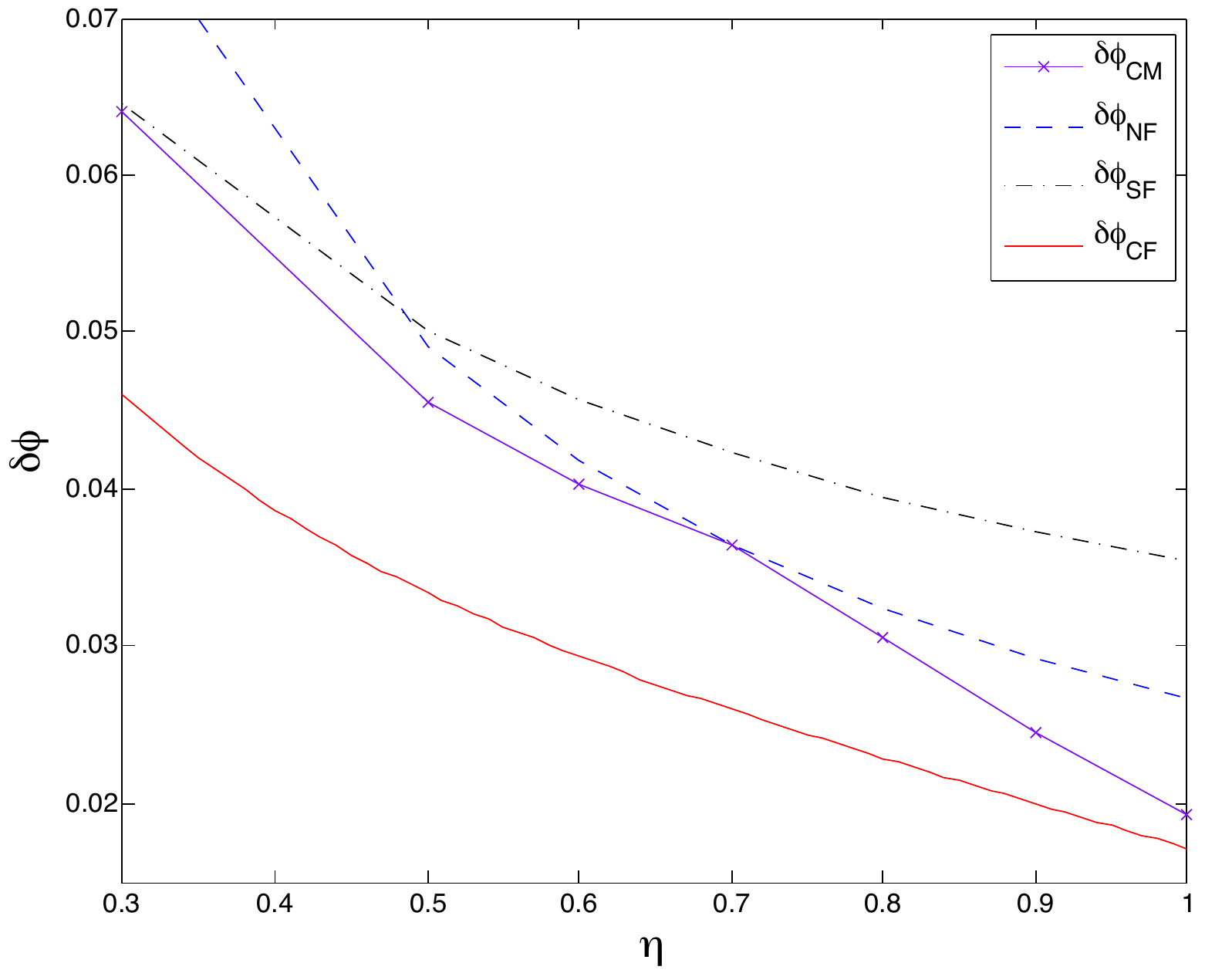, width = 7.5cm}}
  \caption{With a large $\alpha_1$ we can beat both the NOON and SP states. Here, for $\alpha_0=\sqrt{2}$ we beat both NOON and SP all of the time.}
  \label{fig:ECS_larm_vary}
  \vspace{-0.1cm}
\end{figure}

\begin{figure}[!h]
  \vspace{-0.2cm}
  \centering
   {\epsfig{file = 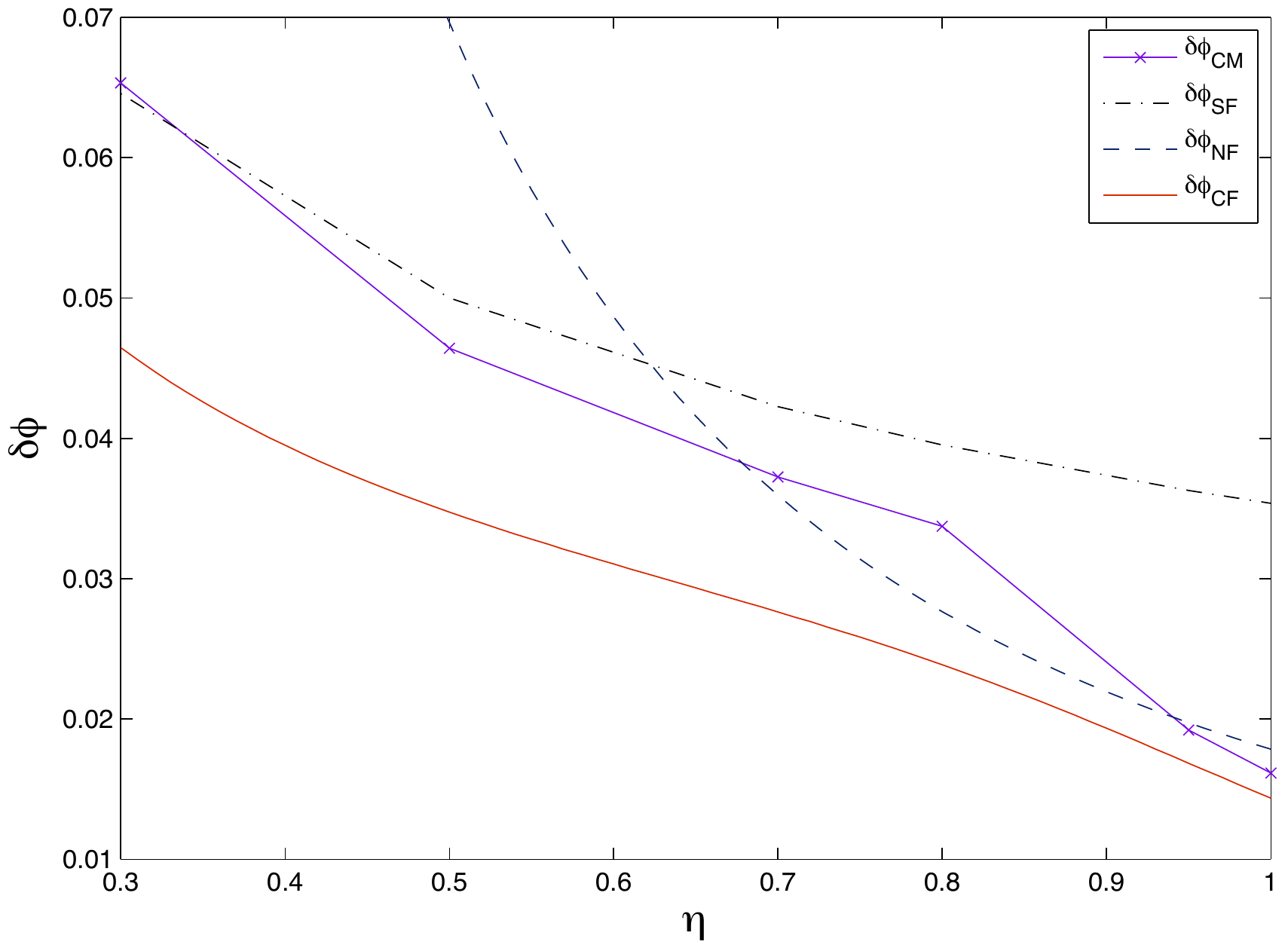, width = 7.5cm}}
  \caption{Here $\alpha=2$. Our scheme beats NOON and SP states for most loss values.}
  \label{fig:ECS_larm_vary2}
  \vspace{-0.1cm}
\end{figure}

\begin{figure}[!h]
  \vspace{-0.2cm}
  \centering
   {\epsfig{file = 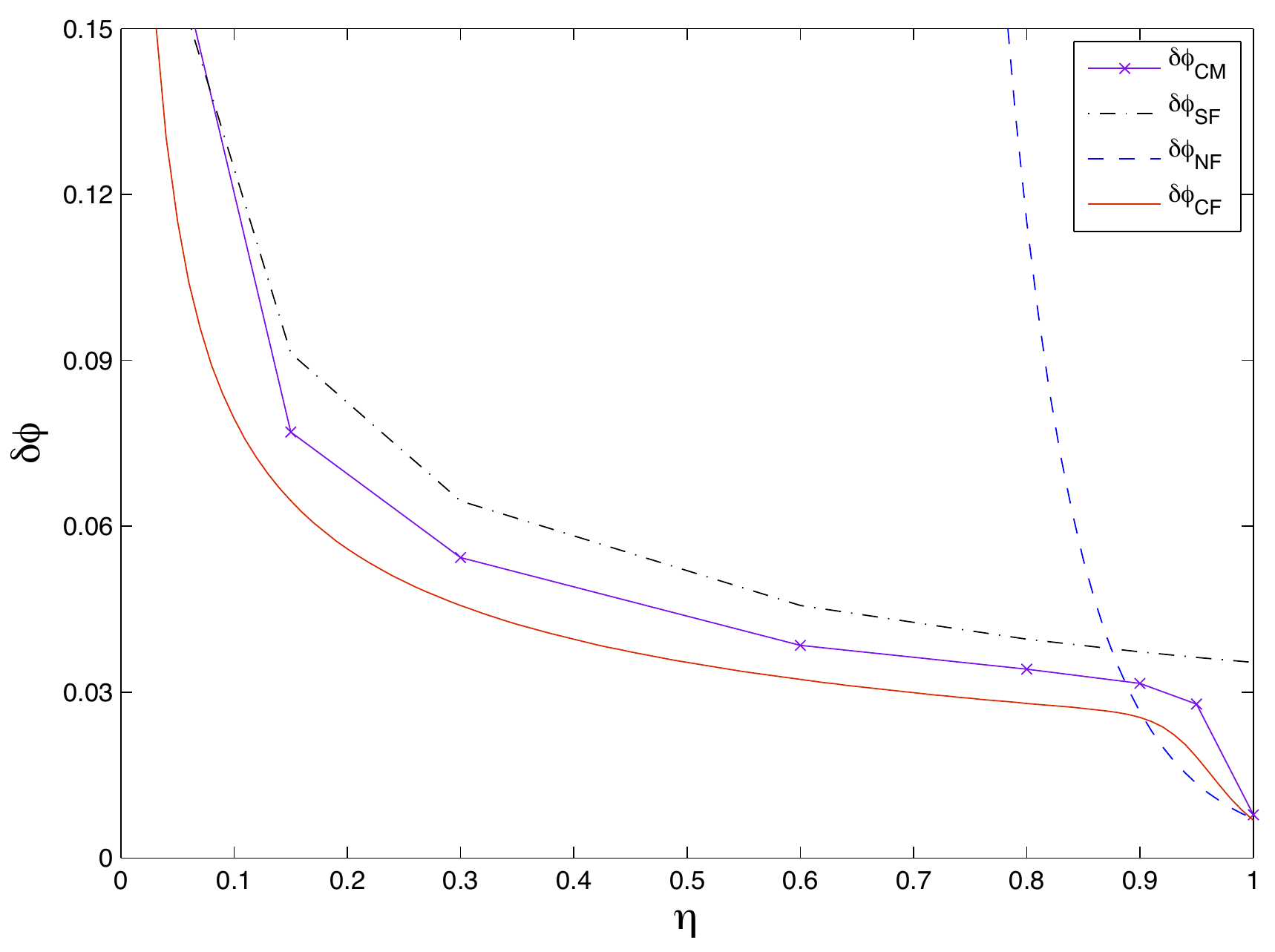, width = 7.5cm}}
  \caption{Here $\alpha=5$. Again we perform better than NOON and SP most of the time.}
  \label{fig:ECS_larm_vary4}
  \vspace{-0.1cm}
\end{figure}

With these changes, and after carefully optimising over $\phi$ and $\alpha_1$, we then obtain the results in Fig.~\ref{fig:ECS_larm_vary} for $\alpha_0=\sqrt{2}$. It can be seen that our state now out-performs the NOON and SP states for all values of loss. Figures \ref{fig:ECS_larm_vary2} and \ref{fig:ECS_larm_vary4} show the results for $\alpha_0=2$ and $\alpha_0=5$ respectively. We can see that for these larger values of $\alpha_0$ our scheme still beats the competitors for the majority of $\eta$ values.

Our results fit well with the Fisher information given by Zhang \textit{et. al.} which is shown as the red solid line $\delta\phi_{CF}$ on all three figures. The authors showed that for large $\alpha$ there is a small region where the NOON state performs better than the ECS because ``although the classical term of the ECS is robust against the photon losses, the Heisenberg term decays about twice as quick as that of the NOON state'' \cite{zhang2013quantum}. This agrees well with our results. Our scheme doesn't saturate the Fisher information, but come reasonably close, and this is enough to beat the NOON and SP states much of the time.

Future work will include examining different ECSs with different QBSs in order to try and come closer to saturating the QFI. We would also like to look how this measurement scheme could be carried out in an experiment. Despite the fact that we have looked at how to \textit{measure} the phase, there are still parts of our scheme that are not easily achievable in experiment, and we would like to iron these parts out so that we have a fully realisable scheme to measure a phase to a significantly higher precision that the competing states.

\section{\uppercase{Conclusions}}
\label{sec:conclusion}

Despite the Fisher information for entangled coherent states showing great potential for robust phase measurement, up to this point it has not been clear how the phase information can actually be measured. Here we show a scheme which can achieve this. When there is no loss our scheme utilises entanglement to perform sub-classical precision. More significantly, we are also able to recapture phase information when there has been loss. This is not possible with single particle or NOON states, and so our results improve upon these competitors for the majority of loss rates. This work brings us ever closer to the ultimate goal in quantum metrology of measuring a phase to a sub-classical precision even when there are significant losses in the system.

\section*{\uppercase{Acknowledgements}}

This work was partly supported by DSTL (contract number DSTLX1000063869).

\vfill
\bibliographystyle{apalike}
{\small
\bibliography{MyLibrary}}

\vfill
\end{document}